\documentclass[prc,twocolumn,aps,showpacs]{revtex4} 
\usepackage{epsf}
\usepackage{amsmath}
\usepackage{amsfonts}
\begin{document}
\title{The role of E1-E2 interplay in 
multiphonon Coulomb excitation}
\author{Alexander Volya} 
\author{Henning Esbensen}
\affiliation{Physics Division, Argonne National Laboratory, Argonne, IL 60439, 
USA}
\begin{abstract}
In this work we study the problem of a charged particle, bound in a 
harmonic-oscillator potential, being excited by the Coulomb 
field from a fast charged 
projectile. Based on a classical solution to the problem and using the 
squeezed-state formalism we are able to treat exactly
both dipole and quadrupole Coulomb field components. 
Addressing various transition amplitudes and 
processes of multiphonon excitation we 
study different aspects resulting from 
the interplay between E1 and E2 fields, ranging from 
classical dynamic polarization effects to questions of quantum interference.
We compare exact calculations with approximate methods.  
Results of this work and the formalism we present can be useful in 
studies of nuclear reaction physics and in atomic stopping theory.  
\end{abstract}
\pacs{25.70.De, 42.50.Dv}
\date{\today}
\maketitle
\section{Introduction}
Recent studies of  
Coulomb-induced breakup of weakly bound nuclei 
indicate a considerable reduction 
of the dissociation probability in comparison with the prediction of 
the first-order Born calculation \cite{esbensen02}. 
The reduction is associated with 
dynamic polarization, where the quadrupole field creates a polarization that
influences the effects induced by the dominating electric dipole. 
This is a well-known phenomenon in atomic 
stopping theory, where the stopping power for charged particles deviates from
the $Z^2$ dependence of Bethe's formula \cite{bethe30}, $Z$ being the charge 
of a particle.
Measurements, as for example, of the
stopping power for protons and anti-protons \cite{andersen89}, 
indicate the presence of a $Z^3$ correction. 
Over the years a substantial progress has been made in measuring 
and understanding of 
this phenomenon
\cite{andersen83,porter83,porter90,novkovic93,pitarke93,arista99,leung89}.  

In this work we consider the 
Coulomb excitation of a charged particle bound by a 
harmonic-oscillator potential. 
The time-dependent electric dipole (E1) 
and quadrupole (E2) fields of a projectile 
excite the particle, and 
within this approximation the 
quantum problem is treated exactly.   
This problem has been studied in the past, 
starting from the classical treatment by Ashley, Ritchie, and Brandt
\cite{ashley72}, followed by quantum perturbation theory 
\cite{hill74}, and later further perturbative Born terms 
\cite{mikkelsen89,mikkelsen90} have also been considered.
Exact numerical solutions to the time-dependent quantum problem
have been demonstrated by Mikkelsen and Flyvbjerg \cite{mikkelsen92}. 
The use of a harmonic oscillator model, as was 
previously argued \cite{mikkelsen89}, is justified for reasons 
of simplicity and at the same time good quality of results in comparison 
with observed stopping powers. 
The preset work adds yet another reason, as it demonstrates how
classical results such as \cite{ashley72,jackson72} can be used  
to obtain exact quantum answers.   
Thus, the approach  presented here 
does not involve heavy numerical calculations.
It presents a simple and transparent way to understand and to treat exactly
the Coulomb excitation of an oscillator.
As an application of the presented technique, rather than repeating previous
numerical calculations,  we concentrate on a less explored 
aspect of Coulomb excitation;   
here the overall excitation probability, the dipole and quadrupole
transition amplitudes, and their interplay 
are discussed.  
  
We use a connection between classical and quantum mechanics 
which is usually complex and 
in most cases takes an approximate quasiclassical form. 
However, in situations where the dynamics of a system is determined by linear 
equations of motion, with  arbitrary time-dependent parameters, 
the classical-to-quantum correspondence is exact. 
The harmonic oscillator which is driven by
the time-dependence of its parameters, such as mass, frequency or external
force, is an example of such a situation.
Most theoretical 
developments in the field of parametric excitations of 
harmonic systems were done more than three decades ago 
\cite{lewis67,lewis69,popov69,popov70,baz}. 
Despite this, it is only
during the past decade theorists began to embrace the beauty of this problem
and to discuss its relevance to many physical processes.
Photon generation, lasers \cite{abdalla98}, quantum interference and 
two-photon excitations \cite{georgiades99}, 
phase transitions, critical phenomena \cite{isar99,drummond01}, 
information theory \cite{aliaga93}, ion traps, 
oriented chiral condensates \cite{volya} 
are examples of recent developments.
Squeezed states that appear as solutions in parametrically driven harmonic
systems, also known as two-photon coherent states,  
generalize the concept of  coherent states
\cite{glauber63} to include two-quanta couplings in the radiation 
field \cite{yuen76,hartley82}.     

By presenting this work we further expand applications 
of the squeezed state formalism 
to problems of Coulomb excitation.
We focus our efforts on classical and quantum effects resulting from the 
interplay of one-(dipole) and two-quanta (quadrupole) 
couplings of an oscillator system to an external field.  
Using symmetries we derive simple formulas for 
excitation probabilities and transition amplitudes that explicitly show
quantum interference effects and can be used to 
improve perturbative calculations.

This work is divided into four sections. In Sec. {\ref{sec:2}} we 
review the properties of harmonic-oscillator systems and the
transition from a classical to a quantum description, which is based on 
canonical transformation. 
We summarize the properties of the parameters that describe the initial state 
to final 
state transition ($S$-matrix) and
derive transition amplitudes and excitation probability relevant 
to Coulomb excitation. 
Section \ref{sec:3} is dedicated to applications. 
Here we consider the classical solution and  address the transition to
the quantum treatment. We discuss 
quantum corrections, perturbative limits, 
charge asymmetry, and two-phonon excitations. The role of E1 and E2 interplay
is emphasized in these discussions.
We summarize our findings in Sec. \ref{sec:4}.

\section{Parametric excitation of coupled oscillator system \label{sec:2}}
We consider a system of $N$ oscillators with the unperturbed Hamiltonian
\begin{equation}
H_{0}=\sum_l \left (\frac{p_l^2}{2}+\frac{\omega_l^2\,q_l^2}{2}\right )\,.
\end{equation} 
These oscillators are disturbed by a time-dependent force $F_l(t)$ and 
time-dependent frequency parameters $Q_{l\,l'}(t)=Q_{l'\,l}(t)\,,$ expressed 
by 
\begin{equation}
H_{\rm int}(t)=
\frac{1}{2}\sum_{l\,l'}\,Q_{l\,l'}(t) q_l q_{l'}+\sum_l F_l(t) q_l\,.
\label{hpert}
\end{equation}
It is assumed that $H_{\rm int}(t)=0$ in the infinite past 
$t\rightarrow -\infty$ and in the infinite future $t\rightarrow \infty\,.$
First we quantize this problem in the infinite past. Here
\begin{subequations}
\begin{eqnarray}
q_l=\frac{1}{\sqrt{2\omega_l}} \left ( a_l e^{-i \omega_l t} + 
a_l^\dagger e^{i \omega_l t}\right ) \,,\\
p_l=-i\sqrt{\frac{\omega_l}{2}} \left ( a_l e^{-i \omega_l t} - 
a^\dagger_l e^{i \omega_l t}\right )\,, 
\end{eqnarray}
\end{subequations}
where $a^\dagger_l$ and $a_l$ are time-independent 
creation and annihilation operators for 
the $l$-th mode.  
For any other time we introduce creation and annihilation operators 
$b^\dagger_l(t)$ and $b_l(t)$
so that the coordinates and momenta in the Heisenberg representation are
\begin{subequations}
\label{qp}
\begin{eqnarray}
q_l=\frac{1}{\sqrt{2\omega_l}} \left ( b_l(t) e^{-i \omega_l t} + 
b^\dagger_l(t) e^{i \omega_l t}\right ) \,,\\
p_l=-i\sqrt{\frac{\omega_l}{2}} \left ( b_l(t) e^{-i \omega_l t} - 
b^\dagger_l(t) e^{i \omega_l t}\right )\,.
\end{eqnarray}
\end{subequations}
The operators  
$b^\dagger(t)$ and $b(t)$ are defined in the interaction representation, as
the
time dependence due to $H_0$ is introduced explicitly in Eq. (\ref{qp}).
For the infinite past we have by definition 
\begin{equation}
b(-\infty)=a\,,\quad b^\dagger(-\infty)=a^\dagger\,. 
\label{bpast}
\end{equation}
At $t\rightarrow +\infty\,$
$b(t)$ and $b^\dagger(t)$ also become time-independent 
\begin{equation}
b(\infty)=b\,,\quad b^\dagger(\infty)=b^\dagger\,.
\end{equation}

The transformation from the 
original operators $a$ and $a^\dagger$ to the intermediate 
$b$ and $b^\dagger$ 
is determined by a general Bogoliubov transformation
\begin{equation}
b_l(t)=\sum_{l'}\left ( u_{l\,l'}(t) a_{l'} + v_{l\,l'}(t) a^\dagger_{l'} 
\right )
+ c_l(t)\,.
\end{equation}
In order to simplify the notation we will write this in matrix form as 
\begin{equation}
b(t)=u(t) a + v(t) a^\dagger+c(t)\,.
\label{bogoliubov}
\end{equation}

The perturbation $
H_{\rm int}$ 
drives the nontrivial time dependence
of these creation and annihilation operators
\begin{equation}
i\frac{d}{dt} b=[b(t),\,H_{\rm int}]={\cal F}(t)+
{\cal Q}^{-}(t) b + 
{\cal Q}^{+}(t) b^\dagger\,,
\label{eqm}
\end{equation}
where  matrices 
\begin{equation}
{{\cal Q}^{\pm}}_{l\,l'}(t)=\frac{Q_{l\,l'}(t)}{2\sqrt{\omega_l\,\omega_{l'}}} 
\,e^{i(\omega_l\pm\omega_{l'})t}\,
\end{equation}
and a vector
$$
{\cal F}_l(t)=\frac{F_l(t)}{\sqrt{2\omega_l}} 
\,e^{i \omega_l t}\,
$$
are introduced. 
The ${{\cal Q}^{-}}$ and ${{\cal Q}^{+}}$ are hermitian and symmetric matrices, 
respectively. 
The time dependence of the parameters $u$ and $v\,,$ as follows from
(\ref{eqm}),  are determined
from 
\begin{subequations}
\label{uv}
\begin{eqnarray}
i\frac{d}{dt}\,u={{\cal Q}^{-}}\, u + {{\cal Q}^{+}}\, v^{*}\,,
\label{uv1}
\\
i\frac{d}{dt}\,v={{\cal Q}^{-}}\, v + {{\cal Q}^{+}}\, u^{*}\,,
\label{uv2}
\end{eqnarray}
\end{subequations}
and
\begin{equation}
i\frac{d}{dt}\,c={{\cal Q}^{-}}\, c + {{\cal Q}^{+}}\, c^{*} + {\cal F}\,.
\label{c}
\end{equation}
These equations are solved numerically with initial conditions  
\begin{equation}
u(-\infty)=1\,,\quad v(-\infty)=0\,,\quad c(-\infty)=0\,,
\end{equation}
which follow from (\ref{bpast}). We stress here that solving  
Eqs. (\ref{uv}) and (\ref{c}) is equivalent to finding a full 
classical solution to the problem.  This follows from the fact that the
equations of motion are linear, so 
the time-dependence of  
a Heisenberg operator becomes identical to the 
time-dependence of its classical analog. 

\subsection{Properties of the $S$-matrix}
In this work we will be interested only in the initial to final state 
transition
probabilities, i.e., in the $S$ matrix that comes from the 
transformation of operators
\begin{equation}
b=u a + v  a^\dagger+c\,,\quad b^\dagger=u^{*} a^\dagger + v^{*} a+c^{*}\,.
\label{uvtrans}
\end{equation}  
The inverse transformation is 
\begin{equation}
a=u^{\dagger} b-v^{T} b^\dagger +c'\,,\quad
a^\dagger= u^{T} b^\dagger-v^{\dagger}+{c'}^{*}\,,
\label{uvinverse}
\end{equation}
where
\begin{equation}
c'=-u^\dagger c + v^{T} c^{*}\,.
\end{equation}
Transformation (\ref{uvtrans}) is canonical and preserves commutation relations
so that if $[a_l,\,a^\dagger_{l'}]=\delta_{l\,l'}$ then 
$[b_l,\,b^\dagger_{l'}]=\delta_{l\,l'}\,;$ this puts some conditions on the 
matrices $u$ and $v$
\begin{subequations}
\label{su}
\begin{eqnarray}
u u^\dagger-v v^\dagger=1\,,\quad u v^{T}-v u^{T}=0\,,
\label{su1} \\
u u^\dagger-v^{T} v^{*}=1\,,\quad u^{\dagger} v-v^{T} u^{*}=0\,.
\label{su2}
\end{eqnarray} 
\end{subequations}

These linear canonical 
transformations, so-called symplectic transformations, 
are often encountered in classical
mechanics. The generators of the above symplectic  
transformation form a group which generalizes the simple $SU(1,1)\,$ group  
\cite{profilo91,gerry87,wei97} of a one-dimensional oscillator. 
In general any infinitesimal time evolution can be described by a linear 
canonical transformation which preserves the phase space in accordance with 
Liouville's theorem.
The beauty of our case is that the linearity remains for
any period of time evolution. 
This allows for an exact classical-to-quantum correspondence. 
Besides the method that we implement in this paper, 
the Wigner transformation \cite{wigner32} that establishes the 
correspondence between
the classical phase space distribution and the quantum density matrix 
can also be used to 
exactly reconstruct the quantum solution from its classical analog
\cite{esbensen81}.
The application of the Wigner transformation in studies of 
various properties
of the harmonic oscillator has been demonstrated repeatedly 
\cite{han88,agarwal91,aliaga93}.

\subsection{Survival probability}
Here we assume that 
$u\,,$ $v$ and $c\,$ are known. Using these we determine
transition amplitudes
between quantum states.
Coherent states,
\begin{equation}
|\alpha \rangle = e^{-|\alpha|^2/2} \sum_{n=0}^{\infty} \frac{\alpha^n}
{\sqrt{n!}} \, |n\rangle \,,
\end{equation}
provide the best means for addressing this problem.
In this basis the evolution operator $U$ is 
\cite{mollow67}
\begin{widetext}
\begin{equation}
\langle \beta |U|\alpha \rangle = A^{0} \exp \left ( 
-\frac{1}{2}(|\beta^2|+|\alpha^2|) +
\frac{1}{2} \beta^{\dagger} v (u^{*})^{-1} \beta^{*}
-\frac{1}{2} \alpha^T (u^{*})^{-1} v \alpha 
+
 \beta^\dagger (u^\dagger)^{-1} \alpha 
-\beta^\dagger (u^{\dagger})^{-1} c' 
- \alpha^{T} (u^{*})^{-1}
c^{*} 
\right )\,.
\label{transitions}
\end{equation}
\end{widetext}
The normalization constant $A^{0}$ is determined by the completeness condition
\begin{equation}
\frac{1}{\pi^N} \int d^{2N} \beta \,|\langle \beta |U|\alpha \rangle|^2 =1\,,
\end{equation}
where integration 
goes through both real and imaginary parts of $\beta\,.$
This is an important parameter, since it sets the scale of all transitions  
and it determines the probability to remain in the ground state
$P_{0, 0}=|\langle 0 |U|0 \rangle|^2={|A^{0}|}^2\,.$
The quantity $P_{0, 0}$  is given by the Gaussian integral, where $\alpha$
is set to zero for simplicity
\begin{equation}
\frac{\pi^N}{P_{0, 0}}=\int d^{2N} \beta \exp \left \{ 
-|\beta|^2+\Re(\beta^{T} \rho \beta
+2c^T \beta- 2c^\dagger \rho \beta) 
\right \}\,.
\label{integral}
\end{equation}

We introduce here the  
matrix 
\begin{equation}
\rho=  v (u^{*})^{-1}\,,
\end{equation}
which is a characteristic of quadrupole excitation. Besides some
trivial normalization that will be further discussed, 
the matrix element $\rho_{i\,j}$ is the 
amplitude of a single step two-quanta excitation of $i$-th and $j$-th modes.
The classical time evolution of this matrix follows from the time
evolution of $u$ and $v$ (\ref{uv}), and is given by 
\begin{equation}
i \frac{d}{dt}\rho=\rho {{\cal Q}^{-}}^{*}  + {{\cal Q}^{-}} \rho + \rho {{\cal Q}^{+}}^{*} \rho 
+ {{\cal Q}^{+}}\,.
\label{eqro}
\end{equation}
This result 
explicitly shows that $\rho$ is a symmetric matrix ($\rho=\rho^{T}$). 

The multidimensional Gaussian integral can be taken as
\begin{equation}
\int d^{2n}{\bf b} \exp (-{\bf b}^{T} {\bf A} {\bf b} + 2{\bf s}^{T} {\bf b}) 
= \frac{\pi^{n}}{\sqrt{\det {\bf A}}} 
\exp\left ( {\bf s}^{T} {\bf A}^{-1} {\bf s} \right )\,,
\label{complexform}
\end{equation}
where relevant to our case 
\begin{equation}
{\bf b}=\left (
\begin{array}{c}
\Re{\beta} \\
\Im{\beta}
\end{array}
\right )\,,\quad
{\bf c}=\left ( \begin{array}{c}
\Re{c} \\
\Im{c}
\end{array} \right )\,,\, 
{\bf A}= \left (
\begin{array}{cc}
1-\Re \rho & \Im \rho \\
\Im \rho & 1+\Re \rho 
\end{array}
\right ),
\end{equation}
\begin{equation}
{\bf s}={\bf Q}^T {\bf c} 
\,,\quad
{\bf Q}=
\left (
\begin{array}{cc}
1-\Re \rho & \Im \rho \\
-\Im \rho & -(1+\Re \rho) 
\end{array}
\right )\,.
\end{equation}
Because of the previously discussed properties of the Bogoliubov 
transformation, inversion of the matrix ${\bf A}$ is not needed. 
It can be shown that 
$ -{\bf c}^{T} {\bf Q} {\bf A}^{-1} {\bf Q}^{T} {\bf c}=
-{c}^{\dagger} {c}+ \Re(c^T \rho^{*} c)\,.$
This results in 
\begin{equation}
P_{0 0}=\sqrt{\det (1-\rho^\dagger \rho)} \,\,\exp 
\left [ -{c}^{\dagger} {c}+ \Re(c^T \rho^{*} c) \right ]\,.
\label{P00}
\end{equation}
The survival probability in the above form unifies the roles of E1 and E2 
processes, and
establishes a direct and simple relation between
the quantities of perturbation theory, $\rho$ and $c\,,$ and the exact answer. 
There is a clear meaning of the different terms in the product (\ref{P00}). 
The $\exp({-c^\dagger c})$ represents the excitation of a coherent state due
to the dipole interaction; the term 
$\sqrt{\det (1-\rho^\dagger \rho)}$ is a result of quadrupole squeezing 
\cite{volya}. The remaining part $\exp[\Re(c^T \rho^{*} c)]$ is a result of
quantum interference.

\subsection{Transition Amplitudes}
Equation (\ref{transitions}) can be used to obtain transition 
amplitudes. Coefficients in a Taylor expansion
of (\ref{transitions}) are related to transition amplitudes 
between oscillator states \cite{volya}, with one-dimensional 
oscillator, for example,
\begin{equation}
\langle \beta|U|\alpha \rangle=  e^{-(|\alpha|^2 + |\beta|^2)/2}\,
\sum_{n, m} \langle n|U|m \rangle \frac{(\beta^*)^n \alpha^m}{\sqrt{n! m!}} \,.
\end{equation}
Below we denote by $| i,j,k \dots\rangle $ the normalized state containing 
single quanta 
in $i,j,k \dots \,$ oscillator modes (later we will use directions 
$x\,$ $y$ and $z$). 
A single-quantum excitation amplitude is given by
\begin{equation}
\langle j |U|0\rangle = - A^0 \{(u^\dagger)^{-1} c'\}_j=
A^0\{(c-\rho c^*)\}_j\,,
\label{AI}
\end{equation}
where the subscript $j$ denotes a component of the vector in the brackets.
Since the interaction vanishes in both the infinite past and infinite future,
the transition probabilities are symmetric with respect to initial and final 
states, in agreement with general
principles of quantum mechanics.
However, the decay amplitude from an excited one-phonon state to the ground
state, which is
\begin{equation}
\langle 0|U| j\rangle = -A^0 \{ (u^*)^{-1} c^*\}_j\,,
\end{equation}
differs from the excitation amplitude by a non trivial 
phase. This emphasizes the effect of quantum 
interference between dipole and quadrupole amplitudes. 
Setting either $c$ or $\rho$ to zero restores the symmetry.

The transition amplitude from the ground state to the excited state with
two oscillator quanta,    
requires expansion of (\ref{transitions}) to the order $\beta^2$
\begin{equation}
\langle i j |U|0\rangle =\frac{ \sqrt{1+\delta_{i j}}}{2} \left (A^0 \rho_{i j}+\frac{
\langle i |U|0\rangle 
\langle j |U|0\rangle }{A^0}\right )\,. 
\label{AII}
\end{equation} 
The first term in the above equation is due to a direct quadrupole transition,
since its amplitude is directly proportional to $\rho\,.$ The second term 
describes the second-order dipole process.

For an initially excited system the transition amplitudes can be calculated in 
a similar manner, by expanding in both $\alpha$ and $\beta\,;$
for example
\begin{equation}
\langle i |U| j\rangle = A^0 (u^\dagger)^{-1}_{i j}+
\frac{
\langle i |U|0\rangle 
\langle 0 |U|j\rangle }{A^0}\,.
\end{equation}

Tailor expansion of a Gaussian form can be used to define 
Hermite polynomials, in particular for the multi-variable case 
such as Eq. (\ref{transitions}). This, in turn, allows for an
analytic expression of transition amplitudes between oscillator states.
Useful iterative relations between amplitudes that follow
from properties of Hermite polynomials can be obtained in this way, 
or alternatively, iterative relations can be derived directly from the
properties of the evolution operator \cite{fernandez89}.

The energy transfer, which is important for calculating the stopping power,
can be calculated using Eq. (\ref{uvtrans}).
For the excitation from the ground state we obtain
\begin{equation}
\Delta E=\langle 0 |\sum_l \omega_l b^{\dagger}_l b_l |0\rangle =
\sum_{l} \omega_l |c_l|^2 +\sum_{l l'} \omega_l |v_{l l'}|^2\,. 
\end{equation}
The above presents an average energy transfer. The energy spread  
can also be calculated 
as in \cite{volya}. 

\subsection{One dimensional harmonic oscillator}
The special case of a one-dimensional harmonic oscillator 
with time-dependent parameters has been extensively studied previously 
\cite{husimi53,popov69,popov70,baz,abdalla86_1,abdalla86_2,abe93}. 
According to Eq. (\ref{P00}) the probability to remain in 
the ground state is
\begin{equation}
P_{0 0}=\sqrt{1-|\rho|^2} \,\,\exp \left\{-|c|^2
\left [1-|\rho|\cos(\phi_{\rho}-2\phi_c)\right ] \right \}\,,
\label{rhot}
\end{equation}
where $\phi_{\rho}$ and $\phi_c$ are phases of $\rho$ and $c\,,$ respectively.
This agrees with the previously obtained result \cite{baz}.
We will further on encounter a one-dimensional case where 
$c=0\,.$ 
Here only even-quanta transitions are possible, and the excitation
probabilities from the ground state and first excited state 
by 
$2n$ quanta are 
\begin{equation}
P_{0, 2n}=\frac{(2n)!}{2^{2n} (n!)^2}\,|\rho|^{2n} \,\sqrt{1-|\rho|^2}\,,
\end{equation}
\begin{equation}
P_{1, 2n+1}=\frac{(2n+1)!}{2^{2n} (n!)^2}\,|\rho|^{2n} \,(1-|\rho|^2)^{3/2}\,.
\end{equation}

\section{Coulomb excitation \label{sec:3}}
We consider here the problem of Coulomb excitation of a particle bound in a 
harmonic-oscillator potential, by a projectile of charge $Z$. We assume that 
the 
projectile moves along a fixed trajectory ${\vec R}(t)\,$
in the $xy$ plane.
In order to utilize the treatment discussed above we expand the Coulomb 
potential up to quadratic terms in the particle coordinate $q\,,$ namely, 
treating it in the dipole-quadrupole approximation
\begin{equation}
V_{\rm coul}(t)=\frac{C}{R^3}\, {\vec q}\cdot 
{\vec R}+\frac{C}{2 R^5}\left ( 
3({\vec q}\cdot {\vec R})^2 - q^2 R^2 \right )\,,
\end{equation}
where $C=Z e^2\,.$ 
The non-zero components of $H_{\rm int}$ defined in Eq. (\ref{hpert}) are 
\begin{equation}
F_x(t)=\frac{C R_x}{R^3}\,,\quad F_y(t)=\frac{C R_y}{R^3}\,,
\end{equation}
$$
Q_{x x}=\frac{C}{R^5}\left ( 
3R_x^2 -  R^2 \right )\,\quad 
Q_{y y}=\frac{C}{R^5}\left ( 
3R_y^2 -  R^2 \right )\,,
$$
\begin{equation}
Q_{x y}=Q_{y x}=\frac{3C R_x R_y}{R^5}\,,\quad Q_{z z}=-\frac{C}{R^3}\,.
\end{equation}

These expressions show that the mode in the direction perpendicular to the
scattering plane is decoupled, and it is excited by the quadrupole term only.
Further, we assume a straight line 
trajectory so that ${\vec R}(t)=(b,\,{\rm v}t,\,0)\,.$ 
This does not simplify the problem, since our formalism 
can be used for any trajectory. This assumption, however, 
allows us to
compare with previous results  \cite{ashley72,hill74,mikkelsen89}.
We introduce the following notations
\begin{equation}
{\mathfrak{d}}_{i}=\frac{C}{b {\rm v} \sqrt {2 \omega_{i}}}\,,\quad {\mathfrak{q}}_{i j}=\frac{C}{2  b^2 {\rm v}\,
\sqrt{\omega_i \omega_j}}\,,
\label{strengths}
\end{equation}
and 
\begin{equation}
\xi_i=\frac{b \omega_i}{{\rm v}}\,,
\end{equation}
where $i$ and $j$ are indices referring to $x\,,$ $y$ and $z\,$ directions.
Often we will assume $\omega_x=\omega_y=\omega_z=\omega\,;$ in that case the 
corresponding indices are omitted.
The parameters ${\mathfrak{d}}$ and ${\mathfrak{q}}$ 
can be viewed as dipole and quadrupole strengths,
so that their ratio $1/(b\sqrt{2\omega})$ determines the applicability of the
multipole expansion.

\subsection{Perturbative treatment}
Here we will discuss several 
approximations. The first one is the 
usual approximation relevant to multipole expansion of the 
Coulomb field. It utilizes the 
smallness of the photon wave length as compared to the size of the system. 
The next two limits
are the ``sudden limit'' where $\xi \ll 1 $ and the ``adiabatic limit''
$\xi \gg 1\,.$ Finally, we will use the term Born approximation which 
corresponds to some order in the iterative solution of the classical 
equations (\ref{uv}) and (\ref{c}). 
Since the classical equations of motion allow here for an exact
reconstruction of the wave function, 
the use 
of this terminology does not contradict its quantum notion.

All of the results in this study, such as in Eqs. (\ref{P00}), (\ref{AI}), and
(\ref{AII}),  are fully determined by the parameters $c$ and 
$\rho\,,$ which may be referred to as shift (dipole) and squeezing 
(quadrupole) parameters, 
respectively.  Both of these parameters can be expanded using the above 
approximations. 
Each order of Born approximation results in the next order 
correction in terms of the quadrupole strength (\ref{strengths}).
Thus an expansion of $c$ and $\rho$ is a series
\begin{equation}
c={\mathfrak{d}} \sum_{n=1,2,\dots} {\mathfrak{c}}^{(n)}(\xi) {\mathfrak{q}}^{n-1}\,,\quad
\rho=  \sum_{n=1,2,\dots} {\mathfrak{r}}^{(n)}(\xi) {\mathfrak{q}}^n\,,
\end{equation}  
where $n=1,2\dots\,$ is the corresponding Born order. 
The coefficients ${\mathfrak{c}}^{(n)}$ and 
${\mathfrak{r}}^{(n)}$ are $\xi$ dependent only.

The lowest order long-range dipole approximation is a  
textbooks example  \cite{bertulani88}. 
Using Eq. (\ref{c}) we obtain
\begin{equation}
c\approx {\mathfrak{d}}\, c^{(1)}=\int_{-\infty}^{\infty}\,-i \,\frac{F_l(t)}{\sqrt{2\omega_l}} 
\,e^{i \omega_l t}\,dt\,.
\end{equation}
This leads to
\begin{equation}
c^{(1)}_x=-2 i  \xi_x K_1\left (\xi_x\right )\,,\quad
c^{(1)}_y=2  \xi_y K_0\left (\xi_y \right )\,,\quad c^{(1)}_z=0\,.
\label{cc}
\end{equation}
We note that $c^{(1)}_y$ is real, and $c^{(1)}_x$ is imaginary. 

Using (\ref{eqro}) the squeezing parameter 
$\rho$ can be calculated in a similar Born-type perturbative manner. 
In the lowest order we obtain 
\begin{subequations}
\label{r}
\begin{eqnarray}
\rho_{x x}= -4i {\mathfrak{q}}_{x x} \,\xi_{x} \left [ K_1(2\xi_x)+2\xi K_0(2\xi_x)\right ]\,,
\label{rxx}
\\
\rho_{y y}= 8i {\mathfrak{q}}_{y y} \,\xi_y^2  K_0(2\xi_y)\,,
\label{ryy}
\\
\rho_{x y}=4 {\mathfrak{q}}_{x y}\,\sqrt{\xi_x \xi_y} \, (\xi_x+\xi_y)\,K_1(\xi_x+\xi_y)\,,
\label{rxy}
\\
\rho_{z z}= 4i {\mathfrak{q}}_{z z} \,\xi_z  K_1(2\xi_z)\,.
\label{rzz}
\end{eqnarray}
\end{subequations}
The above results can also 
be obtained using time dependent perturbation theory.
The first order excitation probability in the $i$-th direction is 
$P^{{\rm I}}_{i}=|c_i|^2\,.$ The
probability to excite quanta $i,j$ in a single step quadrupole process is 
to the 
lowest order 
$P^{{\rm II}}_{i,j}=(1+\delta_{i,j})|\rho_{i j}|^2/4\,.$ These results 
can be compared with the 
exact transition amplitudes in Eqs. (\ref{AI}) and 
(\ref{AII}).

In the sudden limit $\xi\rightarrow 0$ the dipole contribution 
comes from the transverse direction: 
$c_x\approx -2i{\mathfrak{d}}\,,$ while for quadrupole excitations  
$|\rho_{x x}|\approx |\rho_{z z}| \approx 2  {\mathfrak{q}}\,$ are dominant 
in this limit.
The adiabatic limit $\xi\rightarrow \infty$ is a textbook example of 
preservation of adiabatic invariances with exponential precision 
\cite{landau:V1}.

\subsection{$Z^3$ corrections in classical solutions\label{sec:Barkas}}
The long range dipole contribution is the main part of the cross section, 
thus any corrections 
to $c$ are the most important. 
We denote as $c_0(t)$ the inhomogeneous part of the solution to Eq. (\ref{c})
\begin{equation}
c_0(t)=-i \int_{-\infty}^{t} {\cal F}(t') dt' \,. 
\end{equation}
The $c_0\equiv c_0(\infty)=d c^{(1)}$ is the previously discussed first Born
dipole amplitude.
The second Born correction, which here simply means the second order 
approximation to the classical solution of Eq. (\ref{c}) is
\begin{equation}
i\delta c= \int_{-\infty}^\infty [{{\cal Q}^{-}} c_0(t)+{{\cal Q}^{+}} c^{*}_0(t) ]dt\,.
\end{equation}
This integral can be studied numerically or 
perturbatively with respect to $\xi\,.$
\begin{figure}
\begin{center}
\epsfxsize=9.0cm \epsfbox{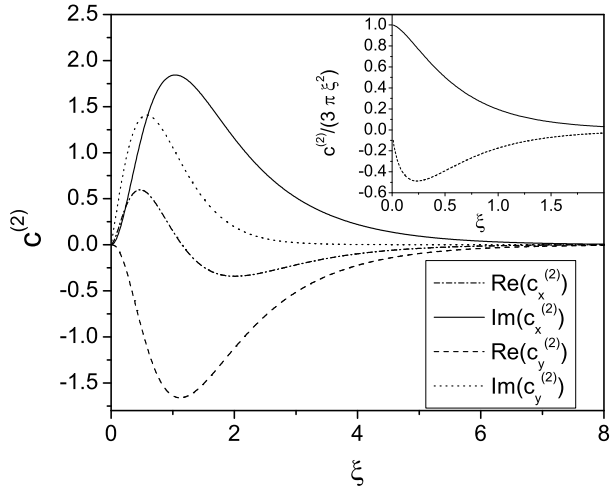}
\end{center}
\caption{The lowest order polarization corrections to the dipole 
amplitude are plotted as a function of $\xi\,.$ The insert 
shows the behavior of $\Im(c^{(2)}_x)$ and $\Re(c^{(2)}_y)$ relative
to $3\pi\xi^2\,.$  \label{corrtoC}
}
\end{figure} 
In Fig. \ref{corrtoC} the dependence of the real and imaginary parts of $c$
are shown as functions of $\xi\,.$ In the adiabatic limit, which corresponds
to large $\xi$ all of these functions decay exponentially:  
$c^{(2)}\sim \exp(-2\xi)$ as dictated by the adiabatic 
nature \cite{landau:V1}.
Approaching the sudden limit all corrections go to zero as powers of $\xi\,.$
In the $\xi\rightarrow 0$ limit 
$\Im{\mathfrak{c}}^{(1)}_x$ is the dominant term, and the lowest order 
correction to it is 
$\Im{\mathfrak{c}}^{(2)}_x\approx 3\pi \xi^2\,.$ 
This leads to the following approximation for the probability
\cite{ashley72}
\begin{equation}
P_{\rm x}=4 {\mathfrak{d}}^2(1-3\pi \mathfrak{q} \xi^2)\,.
\label{barkas}
\end{equation}  
The above correction behaves as $Z^3$ with the charge of the projectile, and
reduces the excitation probability in repulsive Coulomb scattering.
This correction is completely 
classical as it appears in a solution of classical 
differential equations, see \cite{ashley72}. Note that the homogeneous 
part of (\ref{c}) which is the source of this polarization effect, is caused
by the E2 field. 

Higher order corrections to the squeezing parameter $\rho\,$ are relatively 
large but are  generally less important due to the overall smallness of 
the quadrupole amplitude.  Here again we iteratively solve the 
classical equation
(\ref{eqro}). This 
equation  contains no dipole parameters, thus the correction on $\rho$ is
self-induced and would not change in the absence of the E1 field. 
Assuming that
\begin{equation}
i \rho_0(t)= \int_{-\infty}^{t} {{\cal Q}^{+}}(t') dt' =  
\int_{-\infty}^{\infty} dt' 
\Theta(t-t') \,{{\cal Q}^{+}}(t')\,,
\end{equation}
the second Born correction is
\begin{equation}
i \delta \rho=\int_{-\infty}^{\infty} dt \left ( \rho_0(t) {{\cal Q}^{-}}(t)+
{{\cal Q}^{-}}^{*}(t) \rho_0(t) \right )\,.
\end{equation}
For the case of equal frequencies, this expression can be calculated 
analytically 
\begin{subequations}
\begin{eqnarray}
\delta \rho_{x x}=8 {\mathfrak{q}}^2 \xi \left [K_1(2\xi)-2 \xi K_2(2\xi) \right ]
-i \pi {\mathfrak{q}}^2 \xi(1-2 \xi)  e^{-2\xi} \,,\quad
\\
\delta \rho_{x y}=
-2 \pi {\mathfrak{q}}^2 \xi^2  e^{-2\xi}-8 i {\mathfrak{q}}^2 \xi^2 
K_1(2\xi)  \,,\qquad
\\
\delta \rho_{y y}=
- i \pi {\mathfrak{q}}^2 \xi (1+2\xi) e^{-2\xi}\,,\qquad
\\
\delta \rho_{z z}=
-8 {\mathfrak{q}}^2 \xi K_1(2\xi) + \frac{16}{3} i {\mathfrak{q}}^2 \xi^2 K_1(2 \xi)\,.\qquad
\end{eqnarray}
\end{subequations}

In the $\xi\rightarrow 0$ limit, excitations due to the quadrupole field
comes from  
the $x$ and $z$ directions,
leading to the following approximate E2 contribution 
to the excitation probability 
\begin{equation}
\Delta P_{\rho}=
2 {\mathfrak{q}}^2 (1+\pi {\mathfrak{q}} \xi(1- 4\xi)) + 2{\mathfrak{q}}^2 \left 
(1+ \frac{8}{3} {\mathfrak{q}}\xi \right )\,.
\label{deltaro}
\end{equation} 
The charge asymmetry in this case is of opposite sign, resulting in
an enhanced probability for repulsive interaction. This enhancement can 
only be observed in measurements if the 
dominant dipole transitions are blocked or
restricted.

\subsection{Exact treatment}
\begin{figure}
\begin{center}
\epsfxsize=9.0cm \epsfbox{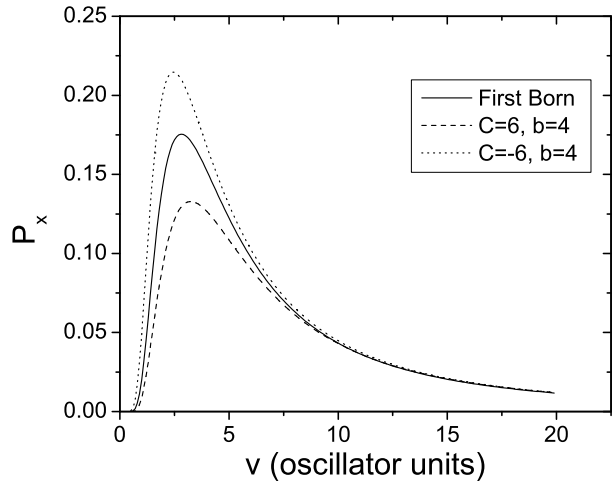}
\end{center}
\caption{Excitation probability as a function of incident velocity, 
at the 
impact parameter $b=4\,.$ The dashed line is the 
result of the repulsive interaction,
$C=6\,.$ The dotted line corresponds to the attractive interaction $C=-6\,.$
These results are compared to the first Born dipole approximation 
shown by the solid line. 
\label{pxc6b4}
}
\end{figure}

We start this section by showing the comparison of the exact excitation 
probability and the first Born dipole 
approximation, $P_{\rm born}=c_0^\dagger c_0\,,$ 
see Fig. \ref{pxc6b4}. 
The harmonic-oscillator model is rather crude and can not be used to fully
understand  features of the realistic Coulomb dissociation of nuclei. However,
it can still provide a qualitative picture.   
For example the Coulomb dissociation of $^{17}_{9}$F can be
looked at assuming that the proton outside the oxygen core 
is bound by a harmonic potential.
The frequency of the oscillator, related to the 
dipole excitation energy,  is assumed to be of the order of
1 MeV. In this case a unit of the oscillator length corresponds to 6.4 fm. 
Assuming Coulomb excitation by $^{56}_{28}$Ni we have $C=6.3\,.$
The choice of parameters used in some 
figures below is guided by this model. In Fig. \ref{pxc6b4} we use
impact parameter $b=4\,$ ($b=25.6$ fm), $C=6\,,$ and the range of 
velocities from 0.2 to 20 (oscillator units) 
that corresponds to the incident energies 
from 0.02 to 200 MeV/u.  The choice of impact parameter is restricted
by the range of applicability of the multipole expansion $\sqrt{2} b \gg 1\,.$ 
Our discussions also imply that the probability to find the nucleon in the 
oscillator at a distance from origin 
larger than impact parameter is negligible.
For $b=2$ this probability is 4\%, assuming the ground state wave 
function. However for $b=4$ that we use, this is less than $10^{-3}$ \%. 

As seen in Fig. \ref{pxc6b4} the effects of higher order corrections
are significant. The a difference between
repulsive and attractive interactions is also transparent. This  
confirms the recent results of more realistic calculations \cite{esbensen02}.
   
\subsubsection{Higher-order corrections}
From the classical solutions  for $\rho$ and $c$ exact quantum results
can be extracted using Eq. (\ref{P00}). 
In order to discuss the impact of a quantum treatment  
we can make an expansion of the exact answer
in Eq. (\ref{P00}), which gives
\begin{equation}
P_{\rm x}=1-P_{0 0}\approx 
c^\dagger c - \frac{(c^\dagger c)^2}{2}+\frac{1}{2} {\rm Tr} (\rho^\dagger \rho) - \Re (c^{T} \rho^* c)\,
\label{apprx}
\end{equation}
$$
=P_{\rm born}+\Delta P_{c}+\Delta P_{\rho}
+\Delta P_{\rm inter}\,.
$$
We identify here three main corrections to the first Born dipole excitation 
probability. 
The first is
$$\Delta P_{c}=c^\dagger c - \frac{(c^\dagger c)^2}{2} - P_{\rm born}\,,$$
which is a correction to the dipole transition probability.   
It is known that the excitation of a quantum oscillator by an external force
creates a coherent state. The dipole amplitude $c\,,$ 
which is related to a classical shift of the oscillator 
from its equilibrium position, appears in the exponent of Eq. (\ref{P00}). 
This accounts for all second and higher order E1
processes. 
Exponentiation of the 
dipole amplitude, in order to account for the loss of strength
to multi-phonon excitations, has been previously discussed  
\cite{norbury93}.  
We stress here that other parts of $\Delta P_{c}\,$ are of classical origin 
and mainly $Z^3$-dependent. They 
come from  the effect of the quadrupole
field on the shift of the oscillator from its equilibrium. This creates
a deviation of $c$ 
from the Born result $c_0$ as discussed in Sec. \ref{sec:Barkas}.

The next term in (\ref{apprx}), 
 $$\Delta P_{\rho}={\rm Tr} (\rho^\dagger \rho) /2\,,$$ 
is a  quadrupole contribution, which comes from
the expansion of the square root in (\ref{P00}). Similar to the 
exponent in the 
dipole term, the use of the square root accounts for all higher order 
quadrupole transitions. 

Finally, a very interesting and purely quantum interference effect 
appears through the
last term in (\ref{apprx}),
$$\Delta P_{\rm inter}=-\Re(c^{T} \rho^* c)\,.$$
Although it may seem that this is a 
quantum Barkas correction, correction vanishes to lowest order. This 
follows from Eqs. (\ref{cc}) and (\ref{r}).

\begin{figure}
\begin{center}
\epsfxsize=9.0cm \epsfbox{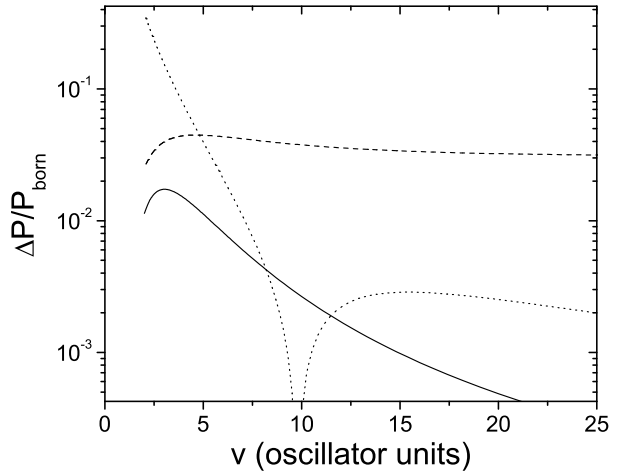}
\end{center}
\caption{
Corrections to the first Born E1 probability. 
Values are relative to $P_{\rm born}$.
Logarithmic scale is used to show contributions of very different magnitude.
Corrections $\Delta P_c/P_{\rm born}$ $\Delta P_\rho/P_{\rm born}$ 
and $\Delta P_{\rm inter}/P_{\rm born}$ are shown in dotted, dashed and solid 
lines, respectively. Parameters $b=4$ and $C=6\,$ are used for this figure.
\label{corr1}
}
\end{figure}

In Fig. \ref{corr1} the absolute values of 
$\Delta P_{c}\,,$ $\Delta P_{\rho}\,,$ and $\Delta P_{\rm inter}$ relative 
to the first Born E1 probability are 
plotted as functions of velocity.  
The figure demonstrates that dominant corrections to the
first order dipole probability come from second and higher 
order dipole transitions, and from the lowest order quadrupole. 
The interference term
contributes only on the level of 
1\%, and quickly vanishes in the sudden limit. 
As discussed above, the interference term vanishes to order $Z^3\,.$ 
Thus it is a $Z^4$ term to leading order. 

\subsubsection{$Z$-dependence}
\begin{figure}
\begin{center}
\epsfxsize=9.0cm \epsfbox{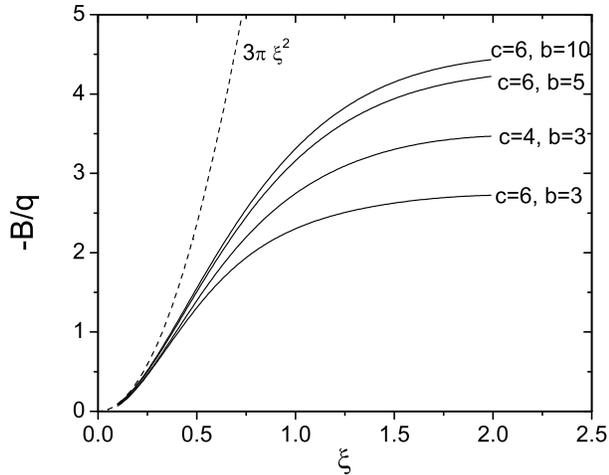}
\end{center}
\caption{The Barkas factor as a function of $\xi\,,$ for various parameters.
The classical correction to the sudden limit, $3\pi\xi^2\,,$ 
is shown by a dashed curve.
\label{brun_x}
}
\end{figure}

In order to examine the $Z$-odd part of the excitation probability from a  
different point of view 
we introduce the Barkas factor
\begin{equation}
B=\frac{P_{\rm x}(Z)-P_{\rm x}(-Z)}{P_{\rm x}(Z)+P_{\rm x}(-Z)}\,.
\end{equation}
As discussed above this effect owes its existence to the classical dynamic
interplay between E1 and E2 fields. 
With the validity of a multipole expansion, the bulk of this
correction is expected to be proportional to the 
quadrupole amplitude ${\mathfrak{q}}\,.$
In Fig. {\ref{brun_x}} $B/{\mathfrak{q}}$ is plotted as a function of $\xi\,.$
Various cases corresponding to different impact parameters and projectile 
charges are considered. In the second Born approximation, the $B/{\mathfrak{q}}$ 
is only a function of $\xi\,,$ thus independent of both charge and impact
parameter \cite{mikkelsen89}. Deviations observed on the plot emphasize the
significance of other corrections that scale as higher powers in ${\mathfrak{q}}$ and 
higher odd powers in $Z\,.$  

\begin{figure}
\begin{center}
\epsfxsize=9.0cm \epsfbox{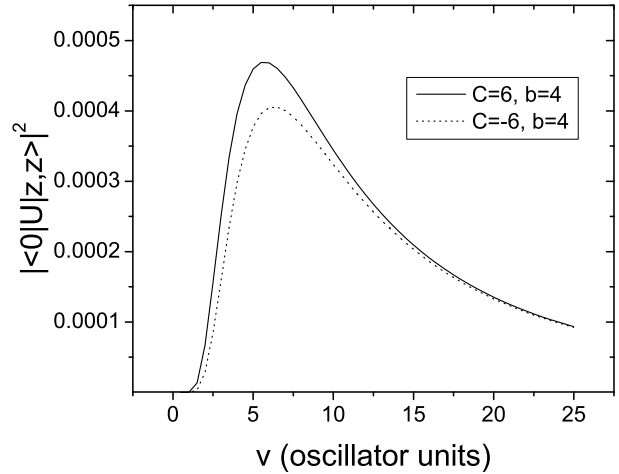}
\end{center}
\caption{The transition probability from the ground state to the two-quantum
excitated state in the 
$z$ direction is shown as a function of incident velocity.
Solid and dotted lines correspond to repulsive and attractive
Coulomb force, respectively. 
\label{qqzz}
}
\end{figure}

The charge asymmetry related to the dipole-quadrupole 
interference appears naturally
in various transition amplitudes. The Fig. \ref{qqzz} shows the
transition probability from the ground state to the state with two quanta
in $z$ direction. In this transition the vanishing out-of-plane dipole force 
makes the quadrupole term dominant. 
As demonstrated by Eq. (\ref{deltaro}) the $Z^3$ Barkas-type charge 
dependence of the quadrupole term $\rho$ is of the opposite sign
in comparison to the dipole-quadrupole case shown in Figs. \ref{pxc6b4} and 
\ref{brun_x}. In agreement with this, an 
enhancement of the excitation probability in repulsive Coulomb interaction
is observed in Fig. \ref{qqzz}.

\subsubsection{Interplay of dipole and quadrupole fields}
\begin{figure}
\begin{center}
\epsfxsize=9.0cm \epsfbox{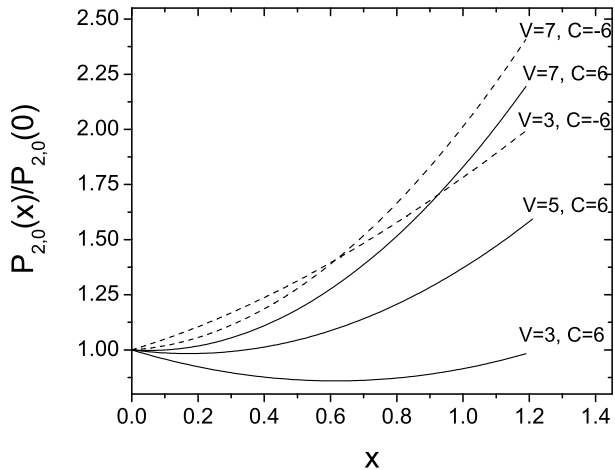}
\end{center}
\caption{Relative two-phonon excitation probability as 
a function of the quadrupole strength. 
\label{twoq}
}
\end{figure}

Our harmonic-oscillator model incorporates E1 and E2 
electromagnetic processes exactly. 
The role and interplay of these
processes in  multi-phonon excitation may be of interest in the 
physics of Giant Dipole Resonances (GDR) 
\cite{bertulani96,norbury93}. Although 
precise calculations with inclusion of the relevant microscopic physics of GDR
are beyond the scope of this paper, we consider here the
effect that the E2 process plays in the two phonon excitation probability
$$
P_{0,2}=\sum_{i<j} |\langle i j|U| 0 \rangle |^2\,.
$$ 
We introduce a quadrupole coupling strength $x$ which simply scales the 
quadrupole part of the perturbation Hamiltonian (\ref{hpert}).
This can be viewed as scaling the quadrupole strength 
${\mathfrak{q}} \rightarrow x {\mathfrak{q}}\,.$ The limits $x=0$ and $x=1$ naturally 
correspond to the cases when E2 is neglected and when E2 is at its 
natural value, respectively.  Eq. (\ref{AII}), as one would expect 
from perturbation theory, shows that the amplitude for a two-phonon excitation
is given by the sum of two terms that reflect the 
first order quadrupole and second order 
dipole processes.
The E1 and E2 interplay is not trivial here and comes in several places.
The first term in  (\ref{AII}) at large quadruple strength is directly 
proportional to $x\,.$ Thus in this limit a quadratic scaling of 
the excitation amplitude is expected. At a smaller strength a substantial
contribution to the second term in (\ref{AII}) appears as a classical 
dynamic polarization effect, influencing the amplitude $c\,.$ This can reduce
the excitation probability for a repulsive interaction. Finally, the 
phases of terms in  
(\ref{AII}) can interfere. It must be noted that the dynamic polarization 
disappears at very low and very high velocities. Also 
both ${\mathfrak{d}}\sim 1/v$ and ${\mathfrak{q}}\sim 1/v$ but
since dipole comes in the second order, the quadrupole amplitude 
dominates at high velocity leading to 
\begin{equation}
P_{0,2}(x)=4 {\mathfrak{q}}^2 x^2\,.
\end{equation}  
In Fig. \ref{twoq} the two-phonon excitation probability as a function
of the coupling strength 
$x$ is shown relative to the case with no E2, $x=0\,.$
Observed features are in agreement with the above discussion.
Cases corresponding to repulsive interaction, shown by solid lines, 
always lead to reduced probability as compared to kinematically 
identical situations but attractive interaction (dashed lines). 
The dynamic polarization effect is maximized around $v=3$ which is consistent
with the enhanced charge asymmetry observed
in Fig. \ref{pxc6b4}.  In this velocity regime, 
the polarization effect is so strong 
that for repulsive interactions ``turning on'' the quadrupole field ($x=1$) 
leads to a reduction of a two-phonon excitation probability. 

\section{Conclusions \label{sec:4}}
In this work we considered a simple model of a charged particle in a 
harmonic-oscillator potential. The system is excited by the electric 
dipole and quadrupole fields of a charged projectile passing by.
We have developed an application of the squeezed-state formalism to 
solve this problem exactly. In terms of the
one-phonon dipole excitation amplitudes $c$ and the 
direct two-phonon quadrupole
excitation 
amplitudes $\rho\,$ we derived simple expressions for various transition
amplitudes, and the total and multi-phonon excitation probabilities.  
The parameters $c$ and $\rho$ are determined from classical equations.
Our results are exact for a harmonic oscillator. 
For a general case  they can be used as a parameterization, extending the
higher-order (multi-phonon) dipole approximation introduced by 
Norbury and Baur 
\cite{norbury93} to include the 
quadrupole field and related interference effects. 

We discussed the interplay between the electric
dipole and quadrupole fields and
the resulting effects on various excitation probabilities.
In the intermediate range of energies, between instant and adiabatic limits,
($0.5<\xi<7$) the influence of dynamic polarization dominates. 
It was shown that the origin of this effect is purely classical.
The quadrupole polarization influences the effect of the dipole force, 
leading to a significant change in the excitation probability. 
This phenomenon, known in stopping power theory as Barkas effect,  
in the lowest order leads to a $Z^3$ charge 
dependence, and reduces the excitation probability for repulsive
Coulomb interactions. We found a similar classical effect of 
self-induced polarization in the quadrupole amplitude. This effect is of the 
opposite sign and for the same repulsive kinematics 
results in an enhancement of the 
quadrupole amplitude. The effect was demonstrated
in $z$-polarized two-phonon excitations.
Our exact results exhibit some additional higher-order 
quantum contributions, but corrections from these are small.
We presented a number of 
numerical calculations and plots that demonstrate a variety of observable
phenomena that can be attributed to the E1-E2 interplay.
 
By presenting this work we hope to introduce the coherent plus squeezed-state
formalism to the field of nuclear reaction physics and nuclear excitations. 
Numerous developments that use dipole excitations 
can be extended to include quadrupole transitions. 
Not every physical situation can be described by a harmonic oscillator.
However, armed with an exact solution
and a full set of time-dependent wave functions,
perturbation theory based on squeezed states
can be a promising future direction.

\begin{acknowledgments}
The authors are thankful to Carlos A. Bertulani for useful discussions. 
This work was supported by the 
U. S. Department of Energy, Nuclear Physics Division, under
contract No. W-31-109-ENG-38.
\end{acknowledgments}

\bibliography{books,coulomb,squeezed}

\begin{thebibliography}{47}
\expandafter\ifx\csname natexlab\endcsname\relax\def\natexlab#1{#1}\fi
\expandafter\ifx\csname bibnamefont\endcsname\relax
  \def\bibnamefont#1{#1}\fi
\expandafter\ifx\csname bibfnamefont\endcsname\relax
  \def\bibfnamefont#1{#1}\fi
\expandafter\ifx\csname citenamefont\endcsname\relax
  \def\citenamefont#1{#1}\fi
\expandafter\ifx\csname url\endcsname\relax
  \def\url#1{\texttt{#1}}\fi
\expandafter\ifx\csname urlprefix\endcsname\relax\def\urlprefix{URL }\fi
\providecommand{\bibinfo}[2]{#2}
\providecommand{\eprint}[2][]{\url{#2}}

\bibitem[{\citenamefont{Esbensen and Bertsch}(2002)}]{esbensen02}
\bibinfo{author}{\bibfnamefont{H.}~\bibnamefont{Esbensen}} \bibnamefont{and}
  \bibinfo{author}{\bibfnamefont{G.~F.} \bibnamefont{Bertsch}},
  \bibinfo{journal}{Nucl. Phys.} \textbf{\bibinfo{volume}{A706}},
  \bibinfo{pages}{477} (\bibinfo{year}{2002}).

\bibitem[{\citenamefont{Bethe}(1930)}]{bethe30}
\bibinfo{author}{\bibfnamefont{H.}~\bibnamefont{Bethe}}, \bibinfo{journal}{Ann.
  d. Phys.} \textbf{\bibinfo{volume}{5}}, \bibinfo{pages}{325}
  (\bibinfo{year}{1930}).

\bibitem[{\citenamefont{Andersen et~al.}(1989)\citenamefont{Andersen,
  Hvelplund, Knudsen, Moller, Pedersen, Uggerhoj, Elsener, and
  Morenzoni}}]{andersen89}
\bibinfo{author}{\bibfnamefont{L.}~\bibnamefont{Andersen}},
  \bibinfo{author}{\bibfnamefont{P.}~\bibnamefont{Hvelplund}},
  \bibinfo{author}{\bibfnamefont{H.}~\bibnamefont{Knudsen}},
  \bibinfo{author}{\bibfnamefont{S.}~\bibnamefont{Moller}},
  \bibinfo{author}{\bibfnamefont{J.}~\bibnamefont{Pedersen}},
  \bibinfo{author}{\bibfnamefont{E.}~\bibnamefont{Uggerhoj}},
  \bibinfo{author}{\bibfnamefont{K.}~\bibnamefont{Elsener}}, \bibnamefont{and}
  \bibinfo{author}{\bibfnamefont{E.}~\bibnamefont{Morenzoni}},
  \bibinfo{journal}{Phys. Rev. Lett.} \textbf{\bibinfo{volume}{62}},
  \bibinfo{pages}{1731} (\bibinfo{year}{1989}).

\bibitem[{\citenamefont{Andersen}(1983)}]{andersen83}
\bibinfo{author}{\bibfnamefont{H.}~\bibnamefont{Andersen}},
  \bibinfo{journal}{Physica Scripta} \textbf{\bibinfo{volume}{28}},
  \bibinfo{pages}{268} (\bibinfo{year}{1983}).

\bibitem[{\citenamefont{Porter and Jeppesen}(1983)}]{porter83}
\bibinfo{author}{\bibfnamefont{L.}~\bibnamefont{Porter}} \bibnamefont{and}
  \bibinfo{author}{\bibfnamefont{R.}~\bibnamefont{Jeppesen}},
  \bibinfo{journal}{Nuclear Instruments and Methods in Physics Research}
  \textbf{\bibinfo{volume}{204}}, \bibinfo{pages}{605} (\bibinfo{year}{1983}).

\bibitem[{\citenamefont{Porter and Lin}(1990)}]{porter90}
\bibinfo{author}{\bibfnamefont{L.}~\bibnamefont{Porter}} \bibnamefont{and}
  \bibinfo{author}{\bibfnamefont{H.}~\bibnamefont{Lin}},
  \bibinfo{journal}{Journal of Applied Physics} \textbf{\bibinfo{volume}{67}},
  \bibinfo{pages}{6613} (\bibinfo{year}{1990}).

\bibitem[{\citenamefont{Novkovic et~al.}(1993)\citenamefont{Novkovic, Subotic,
  Stojanovic, Milosevic, Manic, and Paligoric}}]{novkovic93}
\bibinfo{author}{\bibfnamefont{D.}~\bibnamefont{Novkovic}},
  \bibinfo{author}{\bibfnamefont{K.}~\bibnamefont{Subotic}},
  \bibinfo{author}{\bibfnamefont{M.}~\bibnamefont{Stojanovic}},
  \bibinfo{author}{\bibfnamefont{Z.}~\bibnamefont{Milosevic}},
  \bibinfo{author}{\bibfnamefont{S.}~\bibnamefont{Manic}}, \bibnamefont{and}
  \bibinfo{author}{\bibfnamefont{D.}~\bibnamefont{Paligoric}},
  \bibinfo{journal}{Journal of the Moscow Physical Society}
  \textbf{\bibinfo{volume}{3}}, \bibinfo{pages}{209} (\bibinfo{year}{1993}).

\bibitem[{\citenamefont{Pitarke et~al.}(1993)\citenamefont{Pitarke, Ritchie,
  Echenique, and Zaremba}}]{pitarke93}
\bibinfo{author}{\bibfnamefont{J.}~\bibnamefont{Pitarke}},
  \bibinfo{author}{\bibfnamefont{R.}~\bibnamefont{Ritchie}},
  \bibinfo{author}{\bibfnamefont{P.}~\bibnamefont{Echenique}},
  \bibnamefont{and} \bibinfo{author}{\bibfnamefont{E.}~\bibnamefont{Zaremba}},
  \bibinfo{journal}{Europhysics Lett.} \textbf{\bibinfo{volume}{24}},
  \bibinfo{pages}{613} (\bibinfo{year}{1993}).

\bibitem[{\citenamefont{Arista and Lifschitz}(1999)}]{arista99}
\bibinfo{author}{\bibfnamefont{N.}~\bibnamefont{Arista}} \bibnamefont{and}
  \bibinfo{author}{\bibfnamefont{A.}~\bibnamefont{Lifschitz}},
  \bibinfo{journal}{Phys. Rev. A} \textbf{\bibinfo{volume}{59}},
  \bibinfo{pages}{2719} (\bibinfo{year}{1999}).

\bibitem[{\citenamefont{Leung}(1989)}]{leung89}
\bibinfo{author}{\bibfnamefont{P.}~\bibnamefont{Leung}},
  \bibinfo{journal}{Phys. Rev. A} \textbf{\bibinfo{volume}{40}},
  \bibinfo{pages}{5417} (\bibinfo{year}{1989}).

\bibitem[{\citenamefont{Ashley et~al.}(1972)\citenamefont{Ashley, Ritchie, and
  Brandt}}]{ashley72}
\bibinfo{author}{\bibfnamefont{J.}~\bibnamefont{Ashley}},
  \bibinfo{author}{\bibfnamefont{R.}~\bibnamefont{Ritchie}}, \bibnamefont{and}
  \bibinfo{author}{\bibfnamefont{W.}~\bibnamefont{Brandt}},
  \bibinfo{journal}{Phys. Rev. B} \textbf{\bibinfo{volume}{5}},
  \bibinfo{pages}{2393} (\bibinfo{year}{1972}).

\bibitem[{\citenamefont{Hill and Merzbacher}(1974)}]{hill74}
\bibinfo{author}{\bibfnamefont{K.}~\bibnamefont{Hill}} \bibnamefont{and}
  \bibinfo{author}{\bibfnamefont{E.}~\bibnamefont{Merzbacher}},
  \bibinfo{journal}{Phys. Rev. A} \textbf{\bibinfo{volume}{9}},
  \bibinfo{pages}{156} (\bibinfo{year}{1974}).

\bibitem[{\citenamefont{Mikkelsen and Sigmund}(1989)}]{mikkelsen89}
\bibinfo{author}{\bibfnamefont{H.}~\bibnamefont{Mikkelsen}} \bibnamefont{and}
  \bibinfo{author}{\bibfnamefont{P.}~\bibnamefont{Sigmund}},
  \bibinfo{journal}{Phys. Rev. A} \textbf{\bibinfo{volume}{40}},
  \bibinfo{pages}{101} (\bibinfo{year}{1989}).

\bibitem[{\citenamefont{Mikkelsen and Flyvbjerg}(1990)}]{mikkelsen90}
\bibinfo{author}{\bibfnamefont{H.}~\bibnamefont{Mikkelsen}} \bibnamefont{and}
  \bibinfo{author}{\bibfnamefont{H.}~\bibnamefont{Flyvbjerg}},
  \bibinfo{journal}{Phys. Rev. A} \textbf{\bibinfo{volume}{42}},
  \bibinfo{pages}{3962} (\bibinfo{year}{1990}).

\bibitem[{\citenamefont{Mikkelsen and Flyvbjerg}(1992)}]{mikkelsen92}
\bibinfo{author}{\bibfnamefont{H.}~\bibnamefont{Mikkelsen}} \bibnamefont{and}
  \bibinfo{author}{\bibfnamefont{H.}~\bibnamefont{Flyvbjerg}},
  \bibinfo{journal}{Phys. Rev. A} \textbf{\bibinfo{volume}{45}},
  \bibinfo{pages}{3025} (\bibinfo{year}{1992}).

\bibitem[{\citenamefont{Jackson and McCarthy}(1972)}]{jackson72}
\bibinfo{author}{\bibfnamefont{J.~D.} \bibnamefont{Jackson}} \bibnamefont{and}
  \bibinfo{author}{\bibfnamefont{R.~L.} \bibnamefont{McCarthy}},
  \bibinfo{journal}{Phys. Rev. B} \textbf{\bibinfo{volume}{6}},
  \bibinfo{pages}{4131} (\bibinfo{year}{1972}).

\bibitem[{\citenamefont{Baz et~al.}(1969)\citenamefont{Baz, Zeldovich, and
  Perelomov}}]{baz}
\bibinfo{author}{\bibfnamefont{A.}~\bibnamefont{Baz}},
  \bibinfo{author}{\bibfnamefont{I.}~\bibnamefont{Zeldovich}},
  \bibnamefont{and}
  \bibinfo{author}{\bibfnamefont{A.}~\bibnamefont{Perelomov}},
  \emph{\bibinfo{title}{Scattering, reactions and decay in nonrelativistic
  quantum mechanics. (Rasseyanie, reaktsii i raspady v nerelyativistskoi
  kvantovoi mekhanike)}} (\bibinfo{publisher}{Jerusalem, Israel Program for
  Scientific Translations}, \bibinfo{year}{1969}).

\bibitem[{\citenamefont{Popov and Perelomov}(1969)}]{popov69}
\bibinfo{author}{\bibfnamefont{V.}~\bibnamefont{Popov}} \bibnamefont{and}
  \bibinfo{author}{\bibfnamefont{A.}~\bibnamefont{Perelomov}},
  \bibinfo{journal}{JETP Lett.} \textbf{\bibinfo{volume}{29}},
  \bibinfo{pages}{738} (\bibinfo{year}{1969}).

\bibitem[{\citenamefont{Popov and Perelomov}(1970)}]{popov70}
\bibinfo{author}{\bibfnamefont{V.}~\bibnamefont{Popov}} \bibnamefont{and}
  \bibinfo{author}{\bibfnamefont{A.}~\bibnamefont{Perelomov}},
  \bibinfo{journal}{JETP Lett.} \textbf{\bibinfo{volume}{30}},
  \bibinfo{pages}{910} (\bibinfo{year}{1970}).

\bibitem[{\citenamefont{Lewis}(1967)}]{lewis67}
\bibinfo{author}{\bibfnamefont{H.}~\bibnamefont{Lewis}},
  \bibinfo{journal}{Phys. Rev. Lett.} \textbf{\bibinfo{volume}{18}},
  \bibinfo{pages}{510, 636} (\bibinfo{year}{1967}).

\bibitem[{\citenamefont{Lewis and Riesenfeld}(1969)}]{lewis69}
\bibinfo{author}{\bibfnamefont{H.}~\bibnamefont{Lewis}} \bibnamefont{and}
  \bibinfo{author}{\bibfnamefont{W.}~\bibnamefont{Riesenfeld}},
  \bibinfo{journal}{J. Math. Phys.} \textbf{\bibinfo{volume}{10}},
  \bibinfo{pages}{1458} (\bibinfo{year}{1969}).

\bibitem[{\citenamefont{Abdalla et~al.}(1998)\citenamefont{Abdalla, Ahmed, and
  Al-Homidan}}]{abdalla98}
\bibinfo{author}{\bibfnamefont{M.}~\bibnamefont{Abdalla}},
  \bibinfo{author}{\bibfnamefont{M.}~\bibnamefont{Ahmed}}, \bibnamefont{and}
  \bibinfo{author}{\bibfnamefont{S.}~\bibnamefont{Al-Homidan}},
  \bibinfo{journal}{J. Phys. A} \textbf{\bibinfo{volume}{31}},
  \bibinfo{pages}{3117} (\bibinfo{year}{1998}).

\bibitem[{\citenamefont{Georgiades et~al.}(1999)\citenamefont{Georgiades,
  Polik, and H.J.}}]{georgiades99}
\bibinfo{author}{\bibfnamefont{N.}~\bibnamefont{Georgiades}},
  \bibinfo{author}{\bibfnamefont{E.}~\bibnamefont{Polik}}, \bibnamefont{and}
  \bibinfo{author}{\bibfnamefont{K.}~\bibnamefont{H.J.}},
  \bibinfo{journal}{Phys. Rev. A} \textbf{\bibinfo{volume}{59}},
  \bibinfo{pages}{676} (\bibinfo{year}{1999}).

\bibitem[{\citenamefont{Isar}(1999)}]{isar99}
\bibinfo{author}{\bibfnamefont{A.}~\bibnamefont{Isar}},
  \bibinfo{journal}{Fortschritte der Physik} \textbf{\bibinfo{volume}{47}},
  \bibinfo{pages}{855} (\bibinfo{year}{1999}).

\bibitem[{\citenamefont{Drummond et~al.}(2001)\citenamefont{Drummond,
  Chaturvedi, Dechoum, and Corney}}]{drummond01}
\bibinfo{author}{\bibfnamefont{P.}~\bibnamefont{Drummond}},
  \bibinfo{author}{\bibfnamefont{S.}~\bibnamefont{Chaturvedi}},
  \bibinfo{author}{\bibfnamefont{K.}~\bibnamefont{Dechoum}}, \bibnamefont{and}
  \bibinfo{author}{\bibfnamefont{J.}~\bibnamefont{Corney}}
  (\bibinfo{year}{2001}), vol. \bibinfo{volume}{56A}, p. \bibinfo{pages}{133}.

\bibitem[{\citenamefont{Aliaga et~al.}(1993)\citenamefont{Aliaga, Crespo, and
  Proto}}]{aliaga93}
\bibinfo{author}{\bibfnamefont{J.}~\bibnamefont{Aliaga}},
  \bibinfo{author}{\bibfnamefont{G.}~\bibnamefont{Crespo}}, \bibnamefont{and}
  \bibinfo{author}{\bibfnamefont{A.~N.} \bibnamefont{Proto}},
  \bibinfo{journal}{Phys. Rev. Lett.} \textbf{\bibinfo{volume}{70}},
  \bibinfo{pages}{434} (\bibinfo{year}{1993}).

\bibitem[{\citenamefont{Volya et~al.}(2000)\citenamefont{Volya, Pratt, and
  Zelevinsky}}]{volya}
\bibinfo{author}{\bibfnamefont{A.}~\bibnamefont{Volya}},
  \bibinfo{author}{\bibfnamefont{S.}~\bibnamefont{Pratt}}, \bibnamefont{and}
  \bibinfo{author}{\bibfnamefont{V.}~\bibnamefont{Zelevinsky}},
  \bibinfo{journal}{Nucl. Phys. A} \textbf{\bibinfo{volume}{A671}},
  \bibinfo{pages}{617} (\bibinfo{year}{2000}).

\bibitem[{\citenamefont{Glauber}(1963)}]{glauber63}
\bibinfo{author}{\bibfnamefont{R.}~\bibnamefont{Glauber}},
  \bibinfo{journal}{Phys. Rev.} \textbf{\bibinfo{volume}{131}},
  \bibinfo{pages}{2766} (\bibinfo{year}{1963}).

\bibitem[{\citenamefont{Yuen}(1976)}]{yuen76}
\bibinfo{author}{\bibfnamefont{H.}~\bibnamefont{Yuen}}, \bibinfo{journal}{Phys.
  Rev. A} \textbf{\bibinfo{volume}{13}}, \bibinfo{pages}{2226}
  (\bibinfo{year}{1976}).

\bibitem[{\citenamefont{Hartley and Ray}(1982)}]{hartley82}
\bibinfo{author}{\bibfnamefont{J.}~\bibnamefont{Hartley}} \bibnamefont{and}
  \bibinfo{author}{\bibfnamefont{J.}~\bibnamefont{Ray}},
  \bibinfo{journal}{Phys. Rev. D} \textbf{\bibinfo{volume}{25}},
  \bibinfo{pages}{382} (\bibinfo{year}{1982}).

\bibitem[{\citenamefont{Profilo and Soliana}(1991)}]{profilo91}
\bibinfo{author}{\bibfnamefont{G.}~\bibnamefont{Profilo}} \bibnamefont{and}
  \bibinfo{author}{\bibfnamefont{G.}~\bibnamefont{Soliana}},
  \bibinfo{journal}{Phys. Rev. A} \textbf{\bibinfo{volume}{44}},
  \bibinfo{pages}{2057} (\bibinfo{year}{1991}).

\bibitem[{\citenamefont{Gerry}(1987)}]{gerry87}
\bibinfo{author}{\bibfnamefont{C.}~\bibnamefont{Gerry}},
  \bibinfo{journal}{Phys. Rev. A} \textbf{\bibinfo{volume}{35}},
  \bibinfo{pages}{2146} (\bibinfo{year}{1987}).

\bibitem[{\citenamefont{Lianfu et~al.}(1997)\citenamefont{Lianfu, Shunjin, and
  Quanlin}}]{wei97}
\bibinfo{author}{\bibfnamefont{W.}~\bibnamefont{Lianfu}},
  \bibinfo{author}{\bibfnamefont{W.}~\bibnamefont{Shunjin}}, \bibnamefont{and}
  \bibinfo{author}{\bibfnamefont{J.}~\bibnamefont{Quanlin}},
  \bibinfo{journal}{Zeitschrift fur Physik B (Condensed Matter)}
  \textbf{\bibinfo{volume}{102}}, \bibinfo{pages}{541} (\bibinfo{year}{1997}).

\bibitem[{\citenamefont{Wigner}(1932)}]{wigner32}
\bibinfo{author}{\bibfnamefont{E.}~\bibnamefont{Wigner}},
  \bibinfo{journal}{Phys. Rev.} \textbf{\bibinfo{volume}{40}},
  \bibinfo{pages}{749} (\bibinfo{year}{1932}).

\bibitem[{\citenamefont{Esbensen}(1981)}]{esbensen81}
\bibinfo{author}{\bibfnamefont{H.}~\bibnamefont{Esbensen}}, in
  \emph{\bibinfo{booktitle}{International School of Physics, Enrico Fermi, on
  Nuclear Structure and Heavy Ion Reactions}}, edited by
  \bibinfo{editor}{\bibfnamefont{R.}~\bibnamefont{Broglia}},
  \bibinfo{editor}{\bibfnamefont{C.}~\bibnamefont{Dasso}}, \bibnamefont{and}
  \bibinfo{editor}{\bibfnamefont{R.}~\bibnamefont{Ricci}}
  (\bibinfo{publisher}{Nuovo Cimento}, \bibinfo{year}{1981}), p.
  \bibinfo{pages}{571}.

\bibitem[{\citenamefont{Agarwal and Arun~Kumar}(1991)}]{agarwal91}
\bibinfo{author}{\bibfnamefont{G.}~\bibnamefont{Agarwal}} \bibnamefont{and}
  \bibinfo{author}{\bibfnamefont{S.}~\bibnamefont{Arun~Kumar}},
  \bibinfo{journal}{Phys. Rev. Lett.} \textbf{\bibinfo{volume}{67}},
  \bibinfo{pages}{3665} (\bibinfo{year}{1991}).

\bibitem[{\citenamefont{Han et~al.}(1988)\citenamefont{Han, Kim, and
  Noz}}]{han88}
\bibinfo{author}{\bibfnamefont{D.}~\bibnamefont{Han}},
  \bibinfo{author}{\bibfnamefont{Y.}~\bibnamefont{Kim}}, \bibnamefont{and}
  \bibinfo{author}{\bibfnamefont{M.}~\bibnamefont{Noz}},
  \bibinfo{journal}{Phys. Rev. A} \textbf{\bibinfo{volume}{37}},
  \bibinfo{pages}{807} (\bibinfo{year}{1988}).

\bibitem[{\citenamefont{Mollow}(1967)}]{mollow67}
\bibinfo{author}{\bibfnamefont{B.~R.} \bibnamefont{Mollow}},
  \bibinfo{journal}{Phys. Rev.} \textbf{\bibinfo{volume}{162}},
  \bibinfo{pages}{1256} (\bibinfo{year}{1967}).

\bibitem[{\citenamefont{Fernandez and Tipping}(1989)}]{fernandez89}
\bibinfo{author}{\bibfnamefont{F.}~\bibnamefont{Fernandez}} \bibnamefont{and}
  \bibinfo{author}{\bibfnamefont{R.}~\bibnamefont{Tipping}},
  \bibinfo{journal}{J. Chem. Phys.} \textbf{\bibinfo{volume}{91}},
  \bibinfo{pages}{5505} (\bibinfo{year}{1989}).

\bibitem[{\citenamefont{Sebawe~Abdalla}(1986{\natexlab{a}})}]{abdalla86_1}
\bibinfo{author}{\bibfnamefont{M.}~\bibnamefont{Sebawe~Abdalla}},
  \bibinfo{journal}{Phys. Rev. A} \textbf{\bibinfo{volume}{34}},
  \bibinfo{pages}{4598} (\bibinfo{year}{1986}{\natexlab{a}}).

\bibitem[{\citenamefont{Sebawe~Abdalla}(1986{\natexlab{b}})}]{abdalla86_2}
\bibinfo{author}{\bibfnamefont{M.}~\bibnamefont{Sebawe~Abdalla}},
  \bibinfo{journal}{Phys. Rev. A} \textbf{\bibinfo{volume}{33}},
  \bibinfo{pages}{2870} (\bibinfo{year}{1986}{\natexlab{b}}).

\bibitem[{\citenamefont{Abe and Ehrhardt}(1993)}]{abe93}
\bibinfo{author}{\bibfnamefont{S.}~\bibnamefont{Abe}} \bibnamefont{and}
  \bibinfo{author}{\bibfnamefont{R.}~\bibnamefont{Ehrhardt}},
  \bibinfo{journal}{Phys. Rev. A} \textbf{\bibinfo{volume}{48}},
  \bibinfo{pages}{986} (\bibinfo{year}{1993}).

\bibitem[{\citenamefont{Husimi}(1953)}]{husimi53}
\bibinfo{author}{\bibfnamefont{K.}~\bibnamefont{Husimi}},
  \bibinfo{journal}{Prog. Theor. Phys.} \textbf{\bibinfo{volume}{9}},
  \bibinfo{pages}{381} (\bibinfo{year}{1953}).

\bibitem[{\citenamefont{Bertulani and Baur}(1988)}]{bertulani88}
\bibinfo{author}{\bibfnamefont{C.}~\bibnamefont{Bertulani}} \bibnamefont{and}
  \bibinfo{author}{\bibfnamefont{G.}~\bibnamefont{Baur}},
  \bibinfo{journal}{Phys. Rep.} \textbf{\bibinfo{volume}{163}},
  \bibinfo{pages}{299} (\bibinfo{year}{1988}).

\bibitem[{\citenamefont{Landau and Lifshitz}(1976)}]{landau:V1}
\bibinfo{author}{\bibfnamefont{L.}~\bibnamefont{Landau}} \bibnamefont{and}
  \bibinfo{author}{\bibfnamefont{E.}~\bibnamefont{Lifshitz}},
  \emph{\bibinfo{title}{Mechanics, Course of theoretical physics vol.1}}
  (\bibinfo{publisher}{Oxford ; New York : Pergamon Press},
  \bibinfo{year}{1976}).

\bibitem[{\citenamefont{Norbury and Baur}(1993)}]{norbury93}
\bibinfo{author}{\bibfnamefont{J.~W.} \bibnamefont{Norbury}} \bibnamefont{and}
  \bibinfo{author}{\bibfnamefont{G.}~\bibnamefont{Baur}},
  \bibinfo{journal}{Phys. Rev. C} p. \bibinfo{pages}{1915}
  (\bibinfo{year}{1993}).

\bibitem[{\citenamefont{Bertulani et~al.}(1996)\citenamefont{Bertulani, Canto,
  Hussein, and de~Toledo~Piza}}]{bertulani96}
\bibinfo{author}{\bibfnamefont{C.}~\bibnamefont{Bertulani}},
  \bibinfo{author}{\bibfnamefont{L.}~\bibnamefont{Canto}},
  \bibinfo{author}{\bibfnamefont{M.}~\bibnamefont{Hussein}}, \bibnamefont{and}
  \bibinfo{author}{\bibfnamefont{A.}~\bibnamefont{de~Toledo~Piza}},
  \bibinfo{journal}{Phys. Rev. C} \textbf{\bibinfo{volume}{53}},
  \bibinfo{pages}{334} (\bibinfo{year}{1996}).

\end{thebibliography}
\end{document}